\documentclass[12pt]{iopart}

\usepackage{iopams}
\usepackage{amssymb}
\usepackage{graphicx}

\input{epsf}

\begin{document}

\title{Quantum geometry and quantum algorithms}

\author{S Garnerone$^{\dag,\sharp}$, A Marzuoli$^\ddagger$ and M Rasetti$^{\dag, \sharp}$}

\address{$\dag$Dipartimento di Fisica, Politecnico di Torino, corso Duca degli Abruzzi 24, 10129 Torino (Italy);\\$\sharp$ Institute of Scientific Interchange, Villa Gualino, Viale Settimio Severo 75, 10131 Torino (Italy);\\
$\ddagger$Dipartimento di Fisica Nucleare e Teorica, Universita' degli Studi di Pavia and Istituto Nazionale di Fisica Nucleare, Sezione di Pavia, via A. Bassi 6, 27100 Pavia (Italy)}
\ead{$\dag$ silvano.garnerone@polito.it;\\ $\ddagger$ annalisa.marzuoli@pv.infn.it;\\ $\dag$ mario.rasetti@polito.it}

\begin{abstract}
Motivated by algorithmic problems arising in quantum field theories whose dynamical 
variables are geometric in nature, we provide a quantum algorithm that efficiently 
approximates the colored Jones polynomial. The construction 
is based on the complete solution of Chern-Simons topological quantum field 
theory and its connection to Wess-Zumino-Witten conformal field theory. 
The colored Jones polynomial is expressed as the expectation value of the evolution of the 
$q$-deformed spin-network quantum automaton. A quantum circuit is constructed 
capable of simulating the automaton and hence of computing such expectation value. 
The latter is efficiently approximated using a standard sampling procedure in quantum computation.
\end{abstract}

\maketitle

\section{Introduction}
A new frontier of quantum information 
is the search for algorithms capable of addressing problems in 
low dimensional geometry and topology. The Jones polynomial 
\cite{Jon} characterizes the topology of knots and links (collections 
of circles in 3-space) and is associated with the expectation 
value of a Wilson loop operator in quantum Chern-Simons field theory 
in three dimensions.
The algebraic content of this theory is encoded into a quantum group structure. The Jones polynomial is the link invariant obtained with all the component knots labeled with the fundamental irrep of the quantum deformation of $SU(2)$, denoted in the following by $SU(2)_q$.
Efficient quantum algorithms for approximating 
the Jones polynomial have been recently proposed in \cite{AhJoLa}. In 
\cite{GaMaRa1} we introduced the $q$-deformed spin network automaton model. 
The spin--network quantum simulator model, which essentially encodes 
the (quantum deformed) $SU(2)$ Racah--Wigner tensor algebra, was shown \cite{GaMaRa1} 
to be capable of implementing families of
finite--states and  discrete--time quantum automata 
which accept the language generated by the braid group,
and whose transition amplitudes are indeed colored Jones polynomials. 
The latter are an extension of the Jones polynomial with arbitrary irreps of $SU(2)_q$ 
labeling the component knots. 
In this paper we shall explicitly construct a quantum circuit 
which efficiently simulates the dynamics of these automata 
and hence, if appropriately sampled with a set of 
measurements, approximates the colored Jones polynomial. We shall 
discuss the complexity of the circuit showing that, since the 
time complexity of the spin network automaton is polynomial 
in the size of the input (depending on the index of the braid group and 
on the number of crossings of the knot diagram), the algorithm 
that efficiently simulates the automata also provides an 
efficient estimation of the link invariant.

The paper is organized as follows. In section 2 we briefly review 
the physical setting of quantum geometry and the role 
that in it play the topological invariants. In section 3 we concisely 
describe the structure of the spin network quantum automaton. In section 
4 we provide the details of the quantum algorithm that approximates 
the colored Jones polynomials and of the corresponding quantum circuit. 
In section 5 we provide a few concluding remarks and discuss possible 
future developments and extensions of the methods and concepts 
introduced in the paper.

\section{Quantum geometry and topological invariants}

General relativity -- the prototype of physical theories whose dynamical variable,
the gravitational field, is geometric in nature -- still represents a major improvement
in the `geometrization programme' stated by Klein and Einstein almost
one century ago. These ideas laid dormant long after the birth of quantum
mechanics and quantum field theory. In particular, the quest for a 
quantum gravity theory dates back to the sixties, when Arnowitt,
Deser and Misner \cite{ArDeMi} introduced the so--called $(3+1)$ decomposition of Einstein
field equations, a hamiltonian reformulation of general relativity to be assumed
as the basic ingredient for constructing a canonical quantization scheme for gravity.
We refer the reader to the classical textbooks \cite{AdBaSc,MiThWh} 
for accounts on quantum general relativistic theories up to the seventies.

Nowadays such approach has been almost abandoned in favor of (hopefully) more effective
quantization schemes, but a number of substantial contributions developed in that golden age
keep circulating. A good example is provided by Wheeler's `geometrodynamics',
which embodies the concept of `quantum geometry' of the physical three--dimensional space,
to be thought of as quantum fluctuations of (diffeomorphism classes of) $3$--metrics 
within the `superspace' \cite{Whe}. As we shall see below, $3$--dimensional extended objects
-- more precisely, smooth $3$--manifolds endowed with riemannian or lorentzian metric
tensors -- with their rich geometric structure play a prominent role in 
Chern--Simons quantum field theories and associated statistical field theories. 
Moreover, models of quantum gravity  in three spacetime dimensions represent by themselves
very useful toy models in view of generalizations to the physically significant
$4$--dimensional case.

Euclidean quantum field theory is the quantization procedure of a classical field theory
based on `functional integration', over the space of quantum fluctuations of the
physical fields $\{\phi\}$, of $\exp [-S(\{\phi\})/\hbar]$, where $S(\{\phi\})$ is the classical 
action defined in the Wick--rotated counterpart of Minkowskii spacetime \cite{Wei}.
This approach to quantization
can be related to classical statistical field theory,
and consequently it inherits the language and methods proper of statistical
mechanics (partition functions, phase transitions, etc.).
This latter feature is particularly fruitful if  some kind of
discretization prescription is applied to the classical theory
and suitably extended to the path integrals which turn out to be interpretable as statistical sums or partition functionals.
Indeed, the most successful quantization scheme for general relativity, 
the `sum over histories', was
proposed by Hawking and Hartle \cite{HaHa} 
borrowing techniques from the
euclidean  path integral approach mentioned above. Its  
discretized version, simplicial quantum gravity, relies on Regge's 
discrete reformulation of classical 
general relativity \cite{Reg} and has been  widely addressed in the last two
decades (see {\em e.g.} \cite{AmCaMa,AmDuJo} and
references therein). 

The geometrization programme referred to at the beginning of this section
 was in some sense rephrased
as a `gauge principle' by Yang and Mills in \cite{YaMi}.
Non--Abelian gauge theories interacting with matter fields and their
quantized counterparts still play a central role in the
physics of fundamental interactions, while pure Yang--Mills theories
(classical and quantum) were recognized to encode a number of
interesting  geometric features
(see {\it e.g.} the reviews \cite{Jac, EgHaGi}).

Within the class of quantum Yang--Mills theories we  focus our
attention  on `topological' quantum
field theories (TQFT), formulated in terms of axioms 
by Atiyah in \cite{Ati} (see also \cite{BiBlRa, Qui}). 
Such theories -- quantized through the
path integral prescription starting from a classical Yang--Mills action 
defined on an orientable riemannian $D$--dimensional space(time) -- are characterized by 
gauge invariant partition functions and observables 
(correlation functions) depending only on the
global structure of the space on which the theories live.
The latter geometric functionals  
are computable by standard techniques in quantum field
theory and provide novel representations of `topological invariants'
for $D$--manifolds (and/or 
for particular submanifols embedded in the ambient space) which are of prime interest
both in mathematics and in theoretical physics. All of this enlightens a new
kind of connection between geometry and quantum physics: in TQFT the
physical degrees of freedom of  spacetime
geometry are global and not local. Four 
dimensional Einstein gravity quantized through the euclidean path integral
is not a TQFT, however the role of quantum $3$--geometry is once more
enhanced since  gravity in three spacetime dimensions can be 
reformulated as a gauge theory closely related to the $SU(2)$ 
Chern--Simons TQFT \cite{Wit,Car}.

Without entering into technical details on TQFT in general, let us just recall
some of the basic ingredients of Chern--Simons quantum field theory.

The classical  $SU(2)$ Chern--Simons action for the $3$--sphere $S^3$ 
(the simplest compact, oriented $3$--manifold without boundary) is given by
\begin{equation}\label{CSaction}
k\,S_{CS}(A)=\frac{k}{4\pi}\int_{S^3}\mbox{tr}(AdA+\frac{2}{3}A \wedge A \wedge A)\,,
\end{equation}
where $A$ is the connection 1--form with value in the Lie algebra  
$su(2)$ of the gauge group, $k$ is the coupling constant, $d$ is the exterior
differential, $\wedge$ is the wedge product of
differential forms and the trace is taken over Lie algebra indices.
The partition function of the quantum theory
is obtained from the `path integral' prescription, by integrating the exponential
of $i$ times the classical action (\ref{CSaction}) over the space of gauge--invariant flat
$SU(2)$ connections (the field variables) according to the formal expression 
\begin{equation} \label{CSfunct}
\mathbf{Z}_{\,CS}\,[S^3; k]\;=\;
\int [DA]\,\exp \left\{\frac{i\, k}{4 \pi}\,
S_{CS}\,(A)\,\right\},
\end{equation}
where the coupling constant $k$ is constrained to be 
a positive integer by the gauge--invariant quantization procedure.
The generating functional (\ref{CSfunct}), written for a 
generic compact oriented $3$--manifold 
$\mathcal{M}^3$ with $\partial \mathcal{M}^3 = \emptyset$, is a global 
invariant, namely it depends only on the topological type of  $\mathcal{M}^3$.
This is basically due to the feature that the space of solutions of
quantum CS theory is finite dimensional \cite{Wit}.

The gauge--invariant 
observables in the quantum CS theory are expectation values of Wilson line operators 
associated with oriented knots (or links) embedded
in the $3$--manifold (commonly referred to as Wilson `loop' operators).
Knots and links are `colored' with irreducible representations (irreps) 
of the gauge group $SU(2)$, restricted 
to values ranging over the set $\{0,1/2,1,3/2,\ldots, k/2\}$. Integer $k$ will be 
related to the deformation parameter $q$ in $U_q (su(2))$, the deformed universal
enveloping algebra of $SU(2)$, with $q= \exp(\frac{-2i\pi}{k+2})$.

In particular, the Wilson loop operator associated with a knot
$K$ carrying a spin--$j$ irreducible representation is defined,
for a fixed root of unity $q$, as (the trace of)
the holonomy of the connection 1--form $A$ evaluated along the closed loop
$K$ $\subset S^3$, namely
\begin{equation}\label{WilK}
\mathbf{W}_j\,[K;q]=\tr_j\,P\exp\oint_KA\,,
\end{equation}
where $P$ denotes path ordering.

For a link $L$ made of a collection of knots $\{K_l\, | \, l=1,...,s\}$, each labeled by an
irrep, the expression of the composite Wilson operator reads
\begin{equation}\label{WilL}
\mathbf{W}_{j_1j_2\ldots j_s}\,[L;q]\,=\,\prod_{l=1}^s\;\mathbf{W}_{j_l}\,[K_l;q]\;.
\end{equation}
In the framework of the path integral quantization procedure, 
expectation values of observables are defined as functional
averages weighted with the exponential of the classical action.
In particular, the functional average of the Wilson operator
(\ref{WilL}) is
\begin{equation}\label{Wilexpect}
\mathcal{E}_{j_1...j_s}\,[L;q]\,=\,\frac{
\int [DA]\;\mathbf{W}_{j_1 \ldots j_s}\,[L]\,\exp^{\,\frac{ik}{ 4\pi}
\,S_{CS}\,(A)}} {\int [DA]\;\exp^{\,\frac{ik}{ 4\pi}
S_{CS}\,(A)}},
\end{equation}
where $S_{CS}\,(A)$ is the CS action for the $3$--sphere given in (\ref{CSaction})
and the generating functional in the denominator is usually normalized
to 1.
It can be shown that this expectation value, which essentially\footnote{These 
polynomials are actually invariants of `framed links', see {\em e.g.} \cite{Lic,Gua}.
 The connection
between $\mathcal{E}_{j_1...j_s}\,[L;q]$ and the genuine colored Jones polynomial is 
$J_{j_1...j_s}(L,q)= $ $\{q^{-3\mathit{w}(L)/4}/(q^{1/2}-q^{-1/2})\}$
$\mathcal{E}_{j_1...j_s}\,[L]$, once suitable normalizations for the unknots 
have been chosen. Here $\mathit{w}(L)$ is the writhe associated with the planar 
diagram $D(L)$ of the oriented link $L$, defined as $\mathit{w}(L)=\sum_p \varepsilon (p)$.
The summation runs over the self crossing points of $D(L)$ and $\varepsilon (p)=\pm 1$ 
according to simple combinatorial rules.
The writhe is easily evaluated from the link diagram by simple counting arguments.}
coincides with the colored Jones polynomial \cite{ReTu,KiMe,Lic}, depends only on the isotopy type of the oriented 
link $L$ and on the set of irreps $\{j_1,...,j_s\}$. The original Jones polynomial \cite{Jon}
is recovered when a spin--$\frac{1}{2}$ representation is placed on each link component.
However the colored link invariants are more effective than Jones' in detecting knots,
as discussed in \cite{RaGoKa}.

The colored invariants (\ref{Wilexpect}) are the basic objects that will be addressed for
computational purposes in the rest of this paper. The reader interested in an
account 
of their construction through the quantum group approach
may refer to \cite{GMRlaser}( sect. 3), where the issue of (unitary)
braid group representations is also considered. In the following section
we shall use yet another kind of approach \cite{Kau}, which relies
on the introduction of the boundary Wess--Zumino--Witten conformal field theory
into the Chern--Simons setting. Such approach provides a particularly useful 
presentation of the colored Jones polynomials as expectation values
of unitary braiding operators in WZW theory.

We leave for the concluding remarks at the end of the paper the discussion 
of possible extensions of the quantum algorithm discussed 
in next session to the other hard problems arising in the theory 
of closed (hyperbolic) 3-manifolds.

\section{The spin network quantum automaton}
In the first subsection of this section we shall briefly review automata theory and define basic concepts of formal language theory. Then we describe the model of quantum automaton relevant in the present context: the 
\textit{spin network quantum automaton}, which provides a natural connection between quantum computation and link invariants \cite{GaMaRa1}.
\subsection{Automata theory}
The theory of automata and formal languages addresses in a rigorous way the notions 
of computing machines and computational processes. We review first some of the basic concepts.

If ${\cal A}$ is an alphabet, made of letters, digits or other symbols, and ${\cal A}^*$ denotes
the set of all finite sequences of words over ${\cal A}$, a language 
${\cal L}$ over ${\cal A}$ is a subset of ${\cal A}^*$. 
 The length of the word $w$ is denoted by $|w|$ and $w_i$ is its $i$'th symbol. The concatenation of two words $u,v \in {\cal L}$ is denoted simply  by $uv$. In the fifties Noam 
Chomsky \cite{Cho} introduced a four--level hierarchy describing formal languages according to their structure
(grammar and syntax): regular languages, 
context--free languages, context--sensitive languages and recursively enumerable languages.
The processing of each language is inherently related to  a particular computing model
(see {\em e.g.} \cite{HoUl} for an account on 
formal languages). 
Here we are interested in finite-state automata, 
the machines able to accept regular languages. 

A deterministic finite state automaton consists 
of a finite set of states $S$, an input alphabet ${\cal A}$, 
a transition function $F:S\times {\cal A} \rightarrow {\rm S}$, 
an initial state $s_{in}$ and a set of accepting states 
$S_{acc}\subset S$. The automaton starts in $s_{in}$ and reads an input word $w$ from left to right. At 
the $i$--th step, if the automaton reads the symbol $w_i$, 
then it updates its state to $s'=F(s,w_i)$, where $s$ is the 
state of the automaton reading $w_i$. 
One says that the word has been accepted if the final state reached after 
reading $w$ is in $S_{acc}$.

In the case of a non-deterministic 
finite-state automaton, the transition function is defined 
as a map $F:S \times {\cal A}\rightarrow {\cal P}({\rm S})$, 
where ${\cal P}({\rm S})$ is the power set of ${\rm S}$. After reading a particular 
symbol, the transition can lead to different 
states, according to some assigned probability distribution. 

Generally speaking, quantum finite-state automata 
are obtained from their classical probabilistic counterparts by moving 
from the notion of (classical) probability, associated with transitions, to 
quantum probability amplitudes. Computation takes place inside 
the computational Hilbert space through unitary matrices. 
In the present context we shall confine our attention to the so-called 
measure-once quantum automaton \cite{MoCr}. The latter is a 
5-tuple $M=\left( Q,\Sigma,U,|\mathbf{q}_0 \rangle,|\mathbf{q}_f \rangle \right)$, 
where $Q$ is a finite set 
of quantum states, $\Sigma$ is 
a finite input alphabet with an end--marker symbol $\#$ 
and $U \left( \Sigma \right):Q \rightarrow Q$ is the set of transition 
functions induced by reading $\Sigma$. The probability amplitude 
for the transition from the state $|\mathbf{q} \rangle$ to the state 
$|\mathbf{q}'\rangle$ upon reading the symbol $\sigma \in \Sigma$ is therefore $\langle \textbf{q}|U(\sigma)|\textbf{q}'\rangle$. 
The state $|\mathbf{q}_0\rangle \in Q$ is the initial 
configuration of the system, and $|\mathbf{q}_f\rangle$ 
is an accepting final state. For all 
states and symbols the function $U(\sigma)$ must be 
represented by unitary operators. The end--marker $\#$ is the last symbol of each input word 
and computation terminates after reading it. At the end 
of the computation the configuration of the automaton is measured, 
if it is in an accepting state then the input is accepted, 
otherwise it is rejected. The probability amplitude for the automaton of accepting 
the string $w$ is given by
$
f_M \left( w \right) = \left\langle \mathbf{q}_f  
\right|U(w) \left| \mathbf{q}_0 \right\rangle, U(w) \equiv \, :\!\!\prod_{w_i \in \Sigma} 
U(w_i)\!\!:$ for $w=\,:\!\!\prod_i w_i \!\!:\;$ ($:\! \cdot \!:\;$ 
denotes ordered product, and for $w$ we used the product symbol to denote concatenations). 
The explicit form of $f_M (w)$  defines the language $\mathcal{L}$ 
accepted by that particular automaton.
If $\hat{P}$ denotes the projector over the accepting states, the probability 
for the automaton of accepting the string $w$ is given by 
$
p_M (w) = \Vert  \hat{P} \,|q_w  \rangle  \Vert ^2,
$
where $|q_w \rangle \equiv U(w) | \mathbf{q}_0 \rangle$. 

\subsection{The $q$-deformed spin network automaton}

In this subsection we review briefly the structure 
of the \textit{q-deformed spin network} automaton model, 
first discussed in \cite{GaMaRa1}. This quantum automaton 
is an extension of the spin network model of computation, 
introduced in \cite{MaRa1} and worked out in \cite{MaRa2}, 
constructed on the combinatorics of the Racah--Wigner 
algebra of the quantum group  $SU(2)_q$. The \textit{q-deformed spin network} 
is a model for a quantum automaton capable of processing the braid group language. 
From now on we shall refer to the model simply as \textit{spin network}, 
subsuming the use of the deformed algebra.

\begin{figure}[htbp]
	\centering
		\includegraphics[height=4cm]{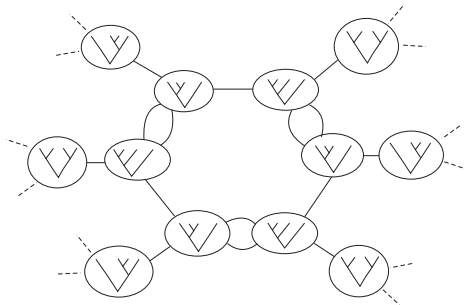}
	\caption{\small{A portion of the spin network graph. 
	Unlabeled trees are associated to particular binary 
	coupling schemes for the total Hilbert space. 
	Single edges correspond to recoupling-like transformations, 
	double edges correspond to braid-like transformations.}}
	\label{fig:qtr2}
\end{figure}

The spin--network can be seen as a collection 
of graphs $\mathfrak{G}_n(V,E)$ parametrized by an integer 
$n$ that is a measure of the size of the automaton. 
For fixed $n$, to each vertex $v \in V$ of $\mathfrak{G}_n(V,E)$ 
is associated the total Hilbert space ${\mathfrak H}^{\textbf{J}}_{\mathfrak{b}}$ 
of the ordered tensor product of $n$ irreps of $SU(2)_q$ (at $q$ root of unity), 
together with a particular binary coupling scheme $\mathfrak{b}$ of the $n$ 
angular momenta $j_l \, (l=1,...,n )\,$ elements of the set $\textbf{J}$. 
Different vertices correspond to different binary coupling schemes and admit 
a realization in terms of unrooted binary trees whose nodes are labeled with $SU(2)_q$ irreps. 
The edges $e \in E$ of $\mathfrak{G}_n(V,E)$ are associated with unitary evolutions 
connecting vertices (Hilbert spaces) belonging to $V$. A restriction is imposed on 
the type of allowed elementary unitary evolutions for the states in ${\mathfrak H}^{\textbf{J}}_{\mathfrak{b}}$: they can be either \textit{braid}-like 
or \textit{recoupling}-like. The former are associated with a unitary 
representation of the braiding between two adjacent leaves of the 
binary tree; the latter are associated with reconfigurations of the 
binary coupling structure of the tree. The graph $\mathfrak{G}_n(V,E)$ 
is constructed in such a way that two vertices are connected by an edge 
if and only if there exist a \textit{braid}-like or a \textit{recoupling}-like 
unitary evolution mapping a state of the first vertex to a state 
of the second vertex (see fig.[\ref{fig:qtr2}]). 

\begin{figure}[htbp]
	\centering
		\includegraphics[height=3cm]{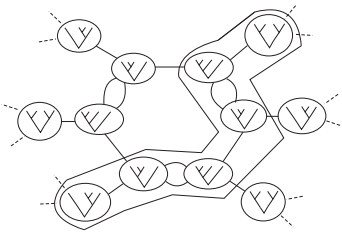}
	\caption{\small{A portion of the spin network graph
	 with a ribbon denoting a path on the graph 
	 corresponding to a particular evolution of the spin-network automaton.}}
	\label{fig:path2}
\end{figure}

\noindent It was shown in \cite{GaMaRa1} 
that it is possible to construct a finite-state quantum 
automaton able to process the language generated by the 
braid group $B_n$. Each graphical realization of the quantum 
automaton can be mapped onto a path in $\mathfrak{G}_n(V,E)$. 
The input-word to the automaton is an element $b \in B_n$ and 
determines the evolution of the automaton according to its image 
$U(b)$, a unitary representation of $B_n$, which constitutes the 
transition rule. The evolution of the automaton is a sequence 
of allowed moves on $\mathfrak{G}_n(V,E)$, as depicted in fig.[\ref{fig:path2}]. 

\noindent The main result in \cite{GaMaRa1} is that 
the probability amplitude for the automaton evolution 
associated to the unitary representation of a braid, 
whose closure is a particular link $L$, is equal to the 
colored Jones polynomial of $L$. This result is based on 
the work of R. Kaul \cite{Kau}. The details of the 
construction of the unitary representations of $B_n$ will be summarized in the next section.

The connection between the spin-network computational model, 
the theory of quantum automata and link invariants will allow 
us to provide a quantum algorithm for the efficient approximation 
of topological invariants of knots. The advantage of using the 
spin network quantum automaton resides in its transition rules, 
which can be straightforwardly expressed in the $q$-deformed 
co-algebra recoupling scheme. The latter naturally provides 
the set of unitary operations which are the building blocks 
of the quantum circuit evaluating the invariants. 
Previous discussion was aimed to mapping the problem of 
evaluating link invariants into the problem of simulating 
the corresponding evolution of the quantum automaton and 
considering henceforth the two problems as equivalent.

\section{A quantum algorithm that approximates the colored Jones polynomial}
In this section we provide a quantum algorithm that efficiently 
approximates the value of the colored Jones polynomial. 
The interest in this problem stems from the fact that an 
additive approximation of the Jones polynomial is sufficient 
to simulate any polynomial quantum computation \cite{BoFrLo}. 
The construction of the algorithm involves three different contexts: 
\begin{enumerate}
 \item a topological context, where the problem is well 
 defined and which allows us to recast the initial instance 
 from the topological language of knot theory to the algebraic language of braid group theory;
 \item a field theoretic context, where tools from CS topological 
 field theory and WZW conformal field theory are used to provide 
 a unitary representation of the braid group;
\item a quantum information context, where the basic features 
of quantum computation are used to efficiently solve the 
original problem formulated in a field theoretic language.
\end{enumerate}

\noindent We shall not discuss the topological context itself, 
where theorems and algorithms are available to relate links and 
braids, and refer the interested reader to \cite{BiBr, Bir} and 
\cite{GMRlaser}. The field theoretic context will be discussed 
in the first subsection. The second subsection will deal with 
the basic structure of the algorithm and its computational 
complexity. In the last subsection we shall complete the proof 
of efficiency and we shall provide notions needed to completely characterize the algorithm.

\subsection{The Kaul construction}
\begin{figure}[htbp]
	\centering
		\includegraphics[height=3cm]{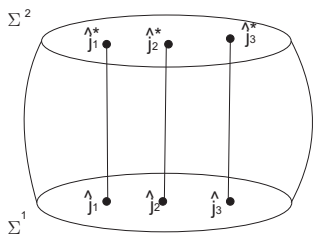}
	\caption{\small{Three Wilson lines intersecting the boundaries of a three-sphere.}}
	\label{fig:punctures}
\end{figure}

In \cite{Kau} R. Kaul provides a unitary representation of the braid group and develops 
a method to evaluate observables in $SU(2)_q$ CS field theory on the 3-sphere $S^3$. 
His construction is based on the relationship between CS 
theory on a three-manifold with boundary and the 
induced WZW conformal field theory on the boundary. 
Let us consider a three-manifold ${\cal M}^3$ with a 
number $n$ of two dimensional boundaries $\Sigma^1, \Sigma^2,...,\Sigma^n$. 
For each of these boundaries, say $\Sigma^i$, there are a number of 
Wilson lines carrying spins $j_l^{\,\,i}$ intersecting 
the boundary at some "`puncture"' $P_l^{\,\,i}$ on the boundary (see fig.[\ref{fig:punctures}]).

We can associate to each $\Sigma^i$ an Hilbert 
space $\mathfrak{H}^i$. The CS functional integral 
over ${\cal M}^3$ is then given as a state in the 
tensor product of such Hilbert spaces. Following the 
literature, in this section we shall henceforth denote 
by $SU(2)_k$ the quantum group $SU(2)_q$ with $q=\exp(\frac{2 \pi i}{k+2})$; 
in the following we shall use both expressions interchangeably. 
The conformal blocks of $SU(2)_k$ WZW field theory on the 
boundaries $\Sigma^i$ with punctures determine the properties 
of $\mathfrak{H}^i$. For each $\mathfrak{H}^i$ there are 
different bases related by duality of the correlators of 
the WZW conformal field theory.

\begin{figure}[htbp]
	\centering
		\includegraphics[height=6cm]{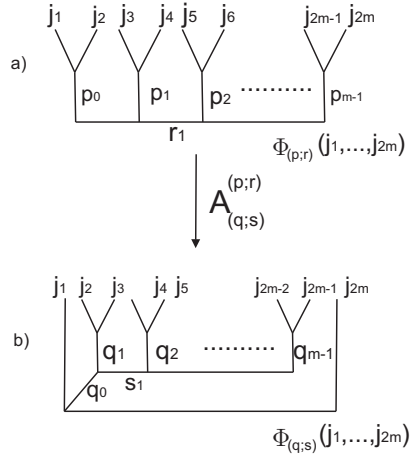}
	\caption{\small{Duality transformation between two types of conformal blocks.}}
	\label{fig:bigduality}
\end{figure}
\noindent These duality matrices can themselves be expressed in terms 
of $q$-deformed $SU(2)$ recoupling coefficients. 
This allows us to relate the unitaries generating the computational 
dynamics of the spin-network automaton to the observables of the CS field theory.

An important aspect in the construction developed in \cite{Kau} is 
the close connection between links and braids. One obtains this important 
result by two main theorems. The first generalizes to colored oriented braids 
a theorem, due to Birman \cite{Bir}, relating links to plats of braids. 
The second, which allows us to decompose the duality matrix associated 
to a general $q$-$3nj$ recoupling transformation into a sequence 
of elementary duality matrices associated to $q$-$6j$ recoupling transformations, reads:

\smallskip

\noindent \textbf{Theorem.} 
\textit{The correlators for 2m primary fields with spins $j_1,j_2,...,j_{2m}$ in $SU(2)_k$ Wess-Zumino-Witten conformal field theory on $S^2$ are related to each other by}

\begin{small}
\begin{equation}
 |\Phi_{\left( \textbf{p};\textbf{r} \right)} \left( j_1,...,j_{2m}\right) \rangle
=
\sum_{\left( \textbf{q};\textbf{s}\right)} A_{\left( \textbf{p};\textbf{r} \right)}^{\left( \textbf{q};\textbf{s}\right)}
\left[
\begin{tabular}{cc}
$j_1$ & $j_2$ \\
$j_3$ & $j_4$ \\
\vdots & \vdots \\
$j_{2m-1}$ & $j_{2m}$\\
\end{tabular} 
\right]
|\Phi_{\left( \textbf{q};\textbf{s} \right)} \left( j_1,...,j_{2m}\right) \rangle,
\end{equation}
\end{small}
\textit{where the duality matrix is given as a product of the basic duality coefficients for the four-point correlators as}
\begin{footnotesize}
\begin{eqnarray}
\noindent
\nonumber
 A_{\left( \textbf{p};\textbf{r}\right) }^{\left( \textbf{q};\textbf{s}\right) }\left [ 
\begin{tabular}{cc}
 $j_1$ & $j_2$ \\
 $\vdots$ & $\vdots$ \\
\end{tabular} 
\right ] 
& = &\sum_{t_1,t_2,...,t_{m-2}} \prod_{i=1}^{m-2} \left( A_{p_i}^{t_i} \left [ 
\begin{tabular}{cc}
$r_{i-1}$ & $j_{2i+1}$ \\
$j_{2i+2}$ & $r_i$ \\
\end{tabular} 
\right ] A_{t_i}^{s_{i-1}}
\left [ 
\begin{tabular}{cc}
$t_{i-1}$ & $q_{i}$ \\
$s_{i}$ & $j_{2m}$ \\
\end{tabular} 
\right ] 
\right ) \\ 
& & \times \prod_{l=0}^{m-2} A_{r_l}^{q_{l+1}}
\left [ 
\begin{tabular}{cc}
$t_{l}$ & $j_{2l+2}$ \\
$j_{2l+3}$ & $t_{l+1}$ \\
\end{tabular} 
\right ]. 
\label{eq:decomposition}
\end{eqnarray} 
\end{footnotesize}

\smallskip

\begin{figure}[htbp]
	\centering
		\includegraphics[height=6cm]{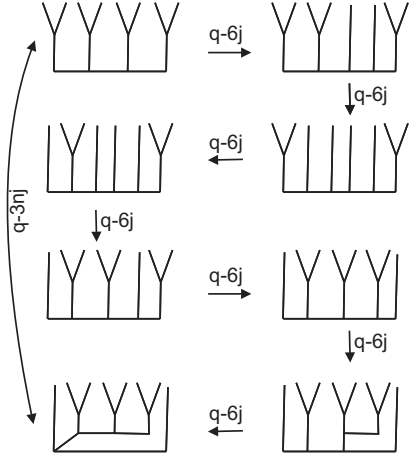}
	\caption{\small{The sequence of decompositions 
	into elementary recoupling transformations of a particular duality matrix.}}
	\label{fig:transformations}
\end{figure}

Here $r_0\equiv p_0,r_{m-2}\equiv p_{m-1},t_0\equiv j_1,t_{m-1}\equiv j_{2m},s_0\equiv q_0,s_{m-2}\equiv
 q_{m-1}, \textbf{j}_{2m}=\sum_{i=1}^{2m-1}\textbf{j}_i$ 
and the spins meeting at trivalent vertices 
in fig.[\ref{fig:bigduality}] satisfy the fusion 
rules of the $SU(2)_k$ CFT. In fig.[\ref{fig:transformations}] 
we provide a pictorial example of the content of the theorem. 

\noindent The elements in the string $\left\lbrace j_1,j_1^*,...,j_m,j_m^* \right\rbrace $ 
will be referred to as $\textbf{j}$-type numbers, the elements in 
$\left\lbrace  p_0,...,p_{m-1} \right\rbrace $ 
as $\textbf{p}$-type numbers and the elements in 
$\left\lbrace  r_0,...,r_{m-2} \right\rbrace $ 
as $\textbf{r}$-type numbers.

\begin{figure}[htbp]
	\centering
		\includegraphics[height=4cm]{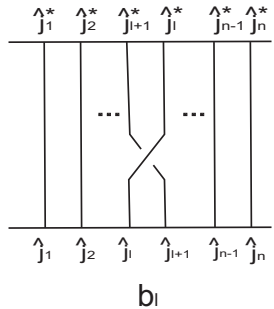}
		\caption{\small{A graphic realization of a generator of the colored braid group.}}
	\label{fig:braidgen}
\end{figure}

A general $n$-strand colored oriented 
braid is specified by giving $n$ assignments 
$\hat{j}_i = \left( j_i, \epsilon_i \right)$, 
representing the spin and the orientation at each 
point on the upper and lower horizontal lines intersecting the strands.
The generators of the groupoid of colored oriented braids are
\begin{equation}\label{eq: braid}
b_l
\left(
\begin{tabular}{cc}
$\hat{j}_{l+1}^*$ & $\hat{j}_{l}^*$\\
$\hat{j}_l$ & $\hat{j}_{l+1}$
\end{tabular}
\right)
\equiv
b_l
\left( 
\begin{tabular}{cccccc}
$\hat{j}_1^*$ & ... & $\hat{j}_{l+1}^*$ & $\hat{j}_{l}^*$ & ... & $\hat{j}_n^*$ \\
$\hat{j}_1$ & ... & $\hat{j}_l$ & $\hat{j}_{l+1}$ & ... & $\hat{j}_n$ \\
\end{tabular} 
\right)
\end{equation}
with $l \in \left\{1,...,n-1\right\}$, where the ``*'' 
implies opposite orientation of the strand with respect to the horizontal line (fig. [\ref{fig:braidgen}]). 

\begin{figure}[htbp]
	\centering
		\includegraphics[height=4cm]{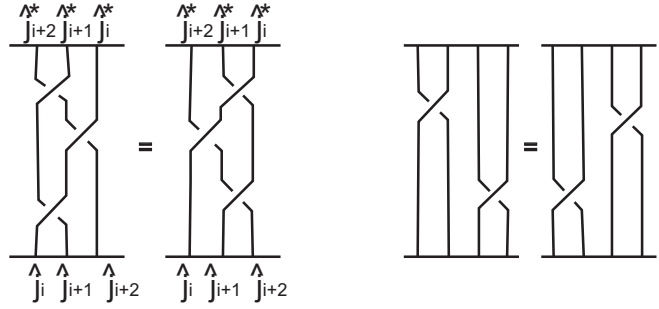}
	\caption{\small{Defining relations for the colored braid group generators.}}
	\label{fig:yangbaxter}
\end{figure}

\noindent The generators of colored oriented braids satisfy the usual defining relations of the braid group (fig.[\ref{fig:yangbaxter}])
$$
b_i \left(
\begin{tabular}{cc}
$\hat{j}_{i+1}^*$ & $\hat{j}_{i}^*$\\
$\hat{j}_i$ & $\hat{j}_{i+1}$
\end{tabular}
\right)
 b_{i+1} \left(
\begin{tabular}{cc}
$\hat{j}_{i+2}^*$ & $\hat{j}_{i}^*$\\
$\hat{j}_{i}$ & $\hat{j}_{i+2}$
\end{tabular}
\right)
 b_i \left(
\begin{tabular}{cc}
$\hat{j}_{i+2}^*$ & $\hat{j}_{i+1}^*$\\
$\hat{j}_{i+1}$ & $\hat{j}_{i+2}$
\end{tabular}
\right)
=
$$
$$
b_{i+1} \left(
\begin{tabular}{cc}
$\hat{j}_{i+2}^*$ & $\hat{j}_{i+1}^*$\\
$\hat{j}_{i+1}$ & $\hat{j}_{i+2}$
\end{tabular}
\right)
 b_i \left(
\begin{tabular}{cc}
$\hat{j}_{i+2}^*$ & $\hat{j}_{i}^*$\\
$\hat{j}_i$ & $\hat{j}_{i+2}$
\end{tabular}
\right)
 b_{i+1} \left(
\begin{tabular}{cc}
$\hat{j}_{i+1}^*$ & $\hat{j}_{i}^*$\\
$\hat{j}_i$ & $\hat{j}_{i+1}$
\end{tabular}
\right)
,
$$
for $i=1,...,n-1$ and
$$
b_i \left(
\begin{tabular}{cc}
$\hat{j}_{i+1}^*$ & $\hat{j}_{i}^*$\\
$\hat{j}_i$ & $\hat{j}_{i+1}$
\end{tabular}
\right)
 b_l \left(
\begin{tabular}{cc}
$\hat{j}_{l+1}^*$ & $\hat{j}_{l}^*$\\
$\hat{j}_l$ & $\hat{j}_{l+1}$
\end{tabular}
\right)
 =b_l \left(
\begin{tabular}{cc}
$\hat{j}_{l+1}^*$ & $\hat{j}_{l}^*$\\
$\hat{j}_l$ & $\hat{j}_{l+1}$
\end{tabular}
\right)
 b_i \left(
\begin{tabular}{cc}
$\hat{j}_{i+1}^*$ & $\hat{j}_{i}^*$\\
$\hat{j}_i$ & $\hat{j}_{i+1}$
\end{tabular}
\right),
$$
for $|i-l|\geq2$.\\

\begin{figure}[htbp]
	\centering
		\includegraphics[height=4cm]{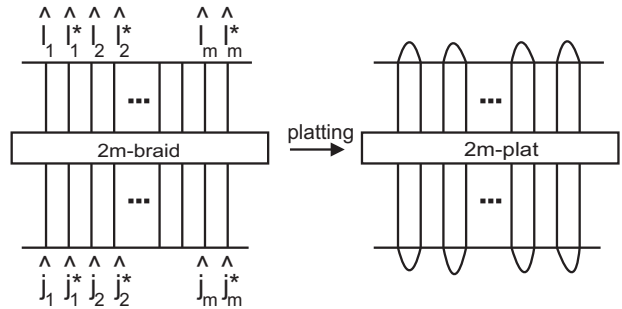}
		\caption{\small{Platting of 2m colored strands.}}
	\label{fig:platting}
\end{figure}

\noindent The \textit{platting} of a colored 
oriented braid on an even number of strands is 
the pairwise joining of contiguous strands, both 
from above and below. Birman's theorem, which relates 
oriented links to plats of ordinary braids \cite{Bir}, 
is extended in \cite{Kau} to colored oriented braids in 
such a way that a colored oriented link is represented by 
the plat closure of an oriented colored braid 
$b\left(
\begin{tabular}{ccccc}
$\hat{l}_1$ & $\hat{l}_1^*$ ... & $\hat{l}_{m}$ & $\hat{l}_{m}^*$ \\
$\hat{j}_1$ & $\hat{j}_1^*$ ... & $\hat{j}_{m}$ & $\hat{j}_{m}^*$ \\
\end{tabular} 
\right)$, see fig.[\ref{fig:platting}].

\begin{figure}[htbp]
	\centering
		\includegraphics[height=4cm]{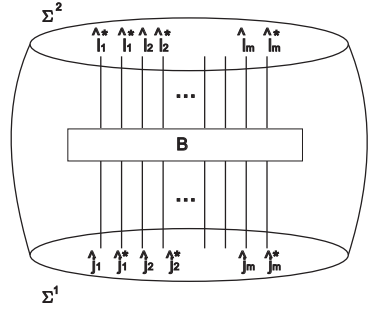}
		\caption{\small{An arbitrary colored braid pattern embedded 
		into a three-manifold with boundaries, $\Sigma^1, \Sigma^2$, with the topology of $S^2$.}}
	\label{fig:wilsonlink}
\end{figure}

\noindent We can finally describe now a 
method for evaluating the expectation value 
of an arbitrary Wilson link operator. Consider the 
three-sphere $S^3$ with two three-balls removed. 
This is a manifold with two boundaries with the topology 
of the 2-sphere $S^2$. Let us place in this manifold $2m$ 
Wilson lines with spins $j_1,j_2,...,j_{2m}$, such that all 
the spins generate an $SU(2)_q$ singlet connecting one boundary 
to the other. It is easily recognized that with these Wilson lines 
we can realize any element of $B_{2m}$ (see fig.[\ref{fig:wilsonlink}]). 

\noindent The CS functional integral over 
the three-manifold can be realized by a state 
in the tensor product of vector spaces $\mathfrak{H}^1 \otimes \mathfrak{H}^2$, 
associated with the two boundaries $\Sigma^1$ and $\Sigma^2$. 
Conformal blocks can be chosen as basis vectors for these vector spaces. 
The inner products of these basis vectors are normalized according to
\begin{equation}
 \langle \Phi_{( \textbf{p};\textbf{r} )} (\hat{j}_1^*,\hat{j}_2^*,...,\hat{j}_{2m}^* )
|\Phi_{( \textbf{u};\textbf{{v}})}(\hat{j}_1,\hat{j}_2,...,\hat{j}_{2m}) \rangle
=\delta_{\textbf{p}, \textbf{u}} \delta_{\textbf{r}, \textbf{v}}.
\end{equation} 

\noindent The basis vectors $| \Phi_{(\textbf{p};\textbf{r})} 
( \hat{j}_1 \hat{j}_2...\hat{j}_{2m}) \rangle$ of the conformal 
blocks $\Phi_{(\textbf{p};\textbf{r})} ( \hat{j}_1 \hat{j}_2...\hat{j}_{2m})$ 
are eigenfunctions of the odd indexed braiding generators $b_{2l+1}$ 
defined in (\ref{eq: braid}). The even indexed braid generators $b_{2l}$ 
are diagonalized in the basis $| \Phi_{(\textbf{q};\textbf{s})} 
( \hat{j}_1,...,\hat{j}_{2m}) \rangle$. The following eigenvalue equations hold
\begin{eqnarray}
\hat{b}_{2l+1} | \Phi_{( \textbf{p};\textbf{r} )} ( \hat{j}_{2l+1}, \hat{j}_{2l+2}) \rangle
&=&\lambda_{p_l} ( \hat{j}_{2l+1}, \hat{j}_{2l+2}) |
\Phi_{( \textbf{p};\textbf{r})}(\hat{j}_{2l+2},\hat{j}_{2l+1})\rangle, \\
\hat{b}_{2l} | \Phi_{( \textbf{q};\textbf{s})} ( \hat{j}_{2l}, \hat{j}_{2l+1})\rangle
&=&\lambda_{q_l} ( \hat{j}_{2l},\hat{j}_{2l+1}) |
\Phi_{( \textbf{q};\textbf{s})}(\hat{j}_{2l+1},\hat{j}_{2l})\rangle.
\end{eqnarray}
Here $| \Phi_{(\textbf{p};\textbf{r})} 
( \hat{j}_{l}, \hat{j}_{l+1}) \rangle \equiv | 
\Phi_{(\textbf{p};\textbf{r})} (\hat{j}_1,..., 
\hat{j}_{l}, \hat{j}_{l+1},...,\hat{j}_{2m}) \rangle$.
The eigenvalues of the braiding matrices depend 
on the relative orientation of the strands and, 
for right-handed half twists (i.e. over-crossings) their value is
\begin{equation}\label{eq: par}
 \lambda_t \left( \hat{j},\hat{i}\right) \equiv \left( - \right)^{j+i-t} 
 q^{\left( c_j+c_i\right)/2 +c_{min\left( i,j\right) } -c_t/2},
\end{equation}
for parallel oriented strands, and
\begin{equation}\label{eq: antipar}
 \lambda_t \left( \hat{j},\hat{i}\right) \equiv \left( - \right)^{|j-i|-t} q^{-|c_j-c_i|/2 +c_t/2},
\end{equation}
if the orientation is anti-parallel. Here $c_j$ is 
the quadratic Casimir operator equal to $j \left( j+1 \right)$ 
for the spin $j$ representation. The eigenvalues (\ref{eq: par}) 
and (\ref{eq: antipar}) derive from the monodromy properties 
of the conformal blocks of the corresponding CFT. The 
associated unitary representation of the braid group is 
provided by the following theorem:

\textbf{Theorem.} \textit{A class of representations $\textbf{K}:B_n 
\rightarrow U(d)$ from the generators of the groupoid of colored 
oriented braids into the unitary $d \times d$ matrices ($d=d(n,|b|)$) 
in the basis} $| \Phi_{\left( \textbf{p};\textbf{r} \right)} \rangle$, \textit{is given by}
\begin{small}
\begin{equation}
\textbf{K}\left[b_{2l+1}  
\left(
\begin{tabular}{cc}
$\hat{j}_{2l+2}^*$ & $\hat{j}_{2l+1}^*$\\
$\hat{j}_{2l+1}$ & $\hat{j}_{2l+2}$\\
\end{tabular} 
\right)
\right]_{\left( \textbf{p};\textbf{r} \right)}^{\left( \textbf{p}';\textbf{r}' \right)}
=\lambda_{p_l} \left( \hat{j}_{2l+1}, \hat{j}_{2l+2} \right) 
\delta_{\textbf{p}}^{\textbf{p}'} \delta_{\textbf{r}}^{\textbf{r}'},
\end{equation}
\end{small}
\textit{and by}
\begin{small}
\begin{eqnarray}\label{eq:evenrep}
\nonumber
\textbf{K}\left[b_{2l} 
\left(
\begin{tabular}{cc}
$\hat{j}_{2l+1}^*$ & $\hat{j}_{2l}^*$\\
$\hat{j}_{2l}$ & $\hat{j}_{2l+1}$\\
\end{tabular} 
\right)
\right]_{\left( \textbf{p};\textbf{r} \right)}^{\left( \textbf{p}';\textbf{r}' \right)}= &
\\
\sum_{\left( \textbf{q};\textbf{s} \right)} 
A_{\left( \textbf{p};\textbf{r} \right)}^{\left( \textbf{q};\textbf{s} \right)}
\left[
\begin{tabular}{cc}
$\vdots$ & $\vdots$ \\
$j_{2l-1}$ & $j_{2l+1}$ \\
$j_{2l}$ & $j_{2l+2}$ \\
$\vdots$ & $\vdots$ \\
\end{tabular}
\right]
\lambda_{q_l} \left( \hat{j}_{2l}, \hat{j}_{2l+1} \right)
A_{\left( \textbf{q};\textbf{s} \right)}^{\left( \textbf{p}';\textbf{r}' \right)}
\left[
\begin{tabular}{cc}
$\vdots$ & $\vdots$ \\
$j_{2l-1}$ & $j_{2l}$ \\
$j_{2l+1}$ & $j_{2l+2}$ \\
$\vdots$ & $\vdots$ \\
\end{tabular}
\right].
\end{eqnarray}
\end{small}
The proof that the defining relations for the braid 
generators are indeed satisfied can be found in \cite{Kau}. 
This result can be used to prove the next\\

\smallskip

\textbf{Theorem.} \textit{The expectation value of a 
Wilson loop operator for an arbitrary link $L$ presented as 
a plat closure of a colored oriented braid 
$
b\left( 
\begin{tabular}{ccccc}
$\hat{l}_1$ & $\hat{l}_1^*$ &...& $\hat{l}_m$ & $\hat{l}_m^*$ \\
$\hat{j}_1$ & $\hat{j}_1^*$ &...& $\hat{j}_m$ & $\hat{j}_m^*$ 
\end{tabular} 
 \right),
$
generated by a word given in terms of the braid generators, is given by}
\small
\begin{eqnarray}
\nonumber
V[L;\textbf{j};q]&=&\prod_{i=1}^{m} \left[ 2j_i+1 \right]
\\
&\times&
\langle \Phi_{(\textbf{0};\textbf{0})} ( \hat{l}_1,...,\hat{l}_m^*)|
\textbf{K} \left[b\left( 
\begin{tabular}{ccc}
$\hat{l}_1$ &...& $\hat{l}_m^*$ \\
$\hat{j}_1$ &...& $\hat{j}_m^*$ 
\end{tabular} 
 \right)
\right]
| \Phi_{(\textbf{0};\textbf{0})}(\hat{j}_1,...,\hat{j}_m^*) \rangle,
\end{eqnarray}
\normalsize
\smallskip
where the multi-index $\left( \textbf{0};\textbf{0} \right)$ 
denotes the case in which all the elements in the set of 
$\textbf{p}$ and $\textbf{r}$ -type numbers are equal to 0, while 
$$
\left[ x \right] \doteq \frac{q^{x/2}-q^{-x/2}}{q^{1/2}-q^{-1/2}}
$$
is the standard notation for the quantum integer. 
The latter theorem gives us the explicit evaluation of 
the colored polynomial. It can be shown that the Jones 
polynomial corresponds to a spin-$\frac{1}{2}$ representation 
living on all the components of the link.

\subsection{The qubit representation}
In this section we show how to efficiently implement 
on a qubit-register the Kaul unitary representation $\textbf{K}$ 
of the colored braid group. We prove that each unitary matrix 
of $\textbf{K} \left( B_{2m} \right)$, interpreted as a gate acting 
on a qubit-register, can be efficiently decomposed into a set of 
universal elementary gates. This is done first encoding into a 
qubit-register the basis vectors used in $\textbf{K}$ and then 
showing how $\textbf{K} \left( b_i \right)$ can be efficiently compiled 
for every $b_i \in B_{2m}$.

\begin{figure}[htbp]
	\centering
		\includegraphics[height=1cm]{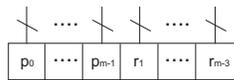}
		\caption{\small{Register of qubits for the Kaul representation.}}
	\label{fig:register}
\end{figure}

Each vector in the basis set 
$\left\lbrace |\Phi_{\left( \textbf{p};\textbf{r} \right)} 
\left( j_1,j_1^*,...,j_m,j_m^* \right) \rangle\right\rbrace $, 
corresponding to the conformal block shown in fig.[\ref{fig:bigduality}a], 
is completely characterized by three sets of quantum numbers, $\textbf{p}$, 
$\textbf{r}$ and $\textbf{j}$, fully labeling the irreps of $SU(2)_q$. 
Recall that the $\textbf{p}$-, $\textbf{r}$- and $\textbf{j}$-type numbers 
belong to the set 
$\left\lbrace  0,\frac{1}{2},...,\frac{k}{2}\right\rbrace$, 
where $k$ is the Chern-Simons coupling constant. 
This means that each type of number can be specified using 
$\lceil \log_2  \left( k+1 \right)  \rceil$ qubits, 
where $\lceil r \rceil$ denotes the least integer $\geq r$. 
An element of the basis can then be encoded 
using $\left( 4m-3 \right) \times \lceil \log_2  \left( k+1 \right)  \rceil$ qubits. 
The register we need to use 
has to code only for the $\textbf{p}$-type and $\textbf{r}$-type 
numbers, implying that only $\left( 2m-3 \right) \times \lceil \log_2  
\left( k+1 \right)  \rceil$ qubits are sufficient.
On the qubit-register we chose the order shown in fig.[\ref{fig:register}]. 

\begin{figure}[htbp]
        \centering
                \includegraphics[height=5cm]{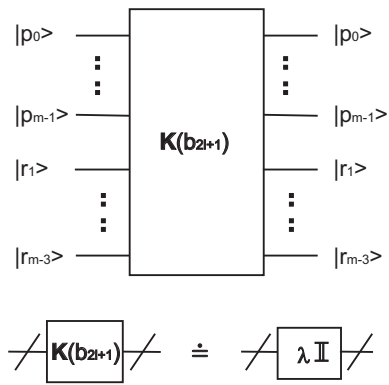}
                \caption{\small{The gate realization of the odd-indexed braid generators.}}
        \label{fig:oddgate}
\end{figure}

\begin{figure}[htbp]
	\centering
		\includegraphics[height=5cm]{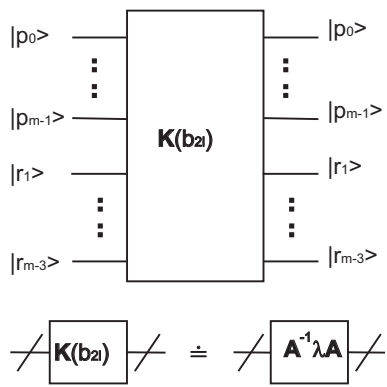}
		\caption{\small{The gate realization of the even-indexed braid generators.}}
	\label{fig:evengate}
\end{figure}

The odd-indexed braid generators are diagonal 
matrices in the basis of the $\textbf{K}$-module, 
therefore there is no problem in implementing their 
action on the quantum register (see fig.[\ref{fig:oddgate}]). 
The even-indexed braid generators have a less trivial 
representation (see fig.[\ref{fig:evengate}]). 

\begin{figure}[htbp]
	\centering
		\includegraphics[height=6cm]{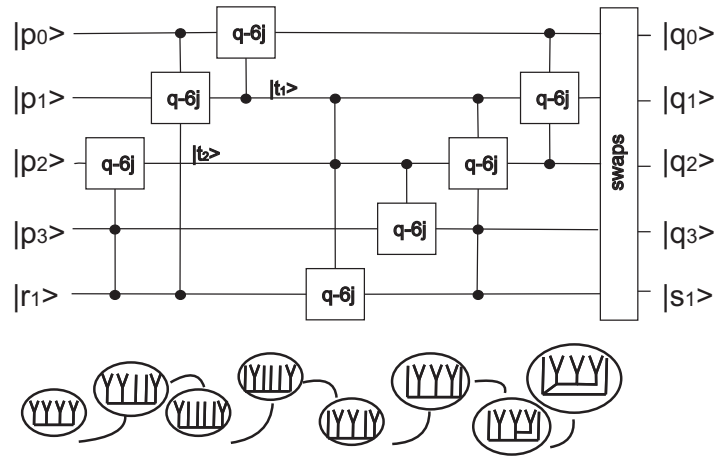}
		\caption{\small{The quantum circuit implementing 
		the decomposition of the eight-point conformal block in 
		terms of q-6j gates. The corresponding path on the spin network graph is shown.}}
	\label{fig:sixpcircuit}
\end{figure}

Resorting to the representation in (\ref{eq:evenrep}) we need to apply two duality 
matrices, or recoupling transformations, in order 
to explicitly construct the image of these generators under $\textbf{K}$. 
Each recoupling transformation can in turn be decomposed 
into a series of elementary \textit{quantum $6j$} transformations 
using (\ref{eq:decomposition}), see e.g. fig.[\ref{fig:sixpcircuit}]. 

\begin{figure}[htbp]
	\centering
		\includegraphics[height=3cm]{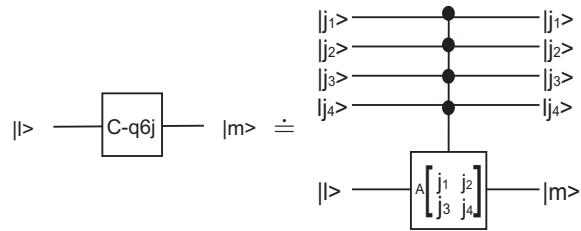}
		\caption{\small{Definition of the controlled q-6j transformation.}}
	\label{fig:defc6j}
\end{figure}

\begin{figure}[htbp]
	\centering
		\includegraphics[height=2cm]{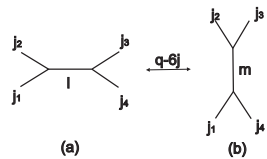}
		\caption{\small{The q-6j transformation.}}
	\label{fig:6j}
\end{figure}

It follows that the problem of efficiently compiling the 
general recoupling transformation from the eigenspace 
of odd-indexed braiding operations to the eigenspace 
of even-indexed braiding operations can be mapped into the easier problem 
of efficiently compiling a single $q$-$6j$ transformation 
(fig.[\ref{fig:defc6j}]). To this end, note that a $q$-$6j$ transformation, 
or the corresponding duality matrix, is a 
unitary transformation from states in the conformal block 
of fig.[\ref{fig:6j}a] into states of the conformal block of fig.[\ref{fig:6j}b].

\noindent Each element of the associated unitary matrix 
is defined in terms of the $q$-Racah coefficients by the following expressions
\begin{eqnarray}\label{eq:duality}
\nonumber
|m \rangle \langle l|_{\textbf{j}} \equiv & A_m^l \left[ 
\begin{tabular}{cc}
$j_1$ & $j_2$ \\
$j_3$ & $j_4$
\end{tabular}
\right] 
=\\
&=\left( - \right)^{\left( j_1+j_2+j_3+j_4 \right)} \sqrt{\left[ 2m+1 \right] \left[ 2l+1 \right]} 
\left( 
\begin{tabular}{ccc}
$j_1$ & $j_2$ & $l$ \\
$j_3$ & $j_4$ & $m$
\end{tabular}
 \right)_q,
\end{eqnarray}
\noindent where all the relevant triplets of $SU(2)_q$-irreps 
satisfy the fusion rules of the WZW CFT. Recall that 
an explicit expression for the $q$-Racah coefficient \cite{Kau} is
\begin{eqnarray}
\nonumber
\left( 
\begin{tabular}{ccc}
$j_1$ & $j_2$ & $l$ \\
$j_3$ & $j_4$ & $m$
\end{tabular}
 \right)_q
=
\Delta \left( j_1,j_2,l \right)
\Delta \left( j_3,j_4,l \right)
\Delta \left( j_1,j_4,m \right)
\Delta \left( j_2,j_3,m \right)
\times
\\
\nonumber
\times
\sum_{x \geq 0} \left( - \right)^x \left[ x+1 \right]!
\left\lbrace
\left[ x-j_1-j_2-l \right]!
\left[ x-j_3-j_4-l \right]!
\left[ x-j_1-j_4-m \right]!
\right. 
\\
\nonumber
\times
\left[ x-j_2-j_3-m \right]!
\left[ j_1+j_2+j_3+j_4-x \right]!
\left[ j_1+j_3+l+m-x \right]!
\\
\times 
\left. 
\left[ j_2+j_4+l+m-x \right]!
\right\rbrace^{-1},
\end{eqnarray}
where 
$
\left[ x \right]! \doteq \left[ x \right] \left[ x-1 \right]!\; {\rm with} \; \left[ 0 \right]!=1,
$
 and $\left[ \cdot \right]$ denotes the q--integer. 
The sum is restricted to all allowed values of $x$ 
such that the quantum integers entering the factorials are non-negative and
$$
\Delta \left( a,b,c \right)=\sqrt{\frac{\left[ -a+b+c \right]! \left[ a-b+c \right]! \left[ a+b-c \right]!}{\left[ a+b+c+1 \right]!}}.
$$
Due to the finiteness of the sum, the coefficients 
(\ref{eq:duality}) can thus be efficiently evaluated classically for 
all the $SU(2)_q$-irreps.

\begin{figure}[htbp]
	\centering
		\includegraphics[height=6cm]{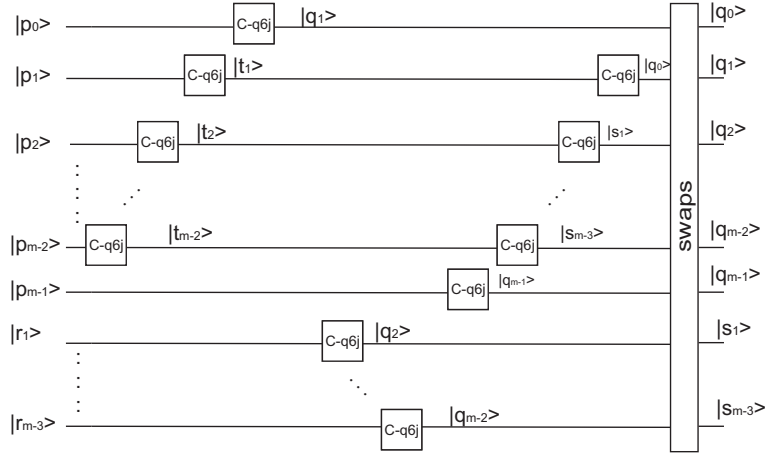}
		\caption{\small{The quantum circuit implementing the general duality transformation.}}
	\label{fig:circuit}
\end{figure}

For what concerns the action on the qubit-register, 
elements (\ref{eq:duality}) belong to unitary matrices of 
rank $ 2^{\lceil \log \left( k+1 \right) \rceil} $, 
parametrized by the set $\textbf{j}$ of those quantum 
numbers which remain unchanged along the transformation. The crucial fact to notice here is that 
the dimension of these matrices is independent of the size of our problem, given by the 
index of the braid group and the number of crossings. 
Since there exist efficient methods to approximate 
unitary matrices of given dimension \cite{HaReCh}, there exists a sequence of universal gates 
that efficiently approximates every $q$-$6j$ transformation. The number of elementary 
$q$-$6j$ transformations needed to decompose a general $q$-$3nj$ recoupling transformation 
is $2m-3$, linear in the size of the problem. In conclusion, 
the Kaul representation $\textbf{K}$ associated with an arbitrary 
colored oriented braid can indeed be efficiently compiled on a 
standard quantum computer. The circuit 
implementing the decomposition of $\textbf{K}$ is shown in fig.[\ref{fig:circuit}].

\subsection{The algorithm}
The general structure of the quantum automaton whose 
dynamical evolution derived in \cite{GaMaRa1} is characterized 
by probability amplitudes whose values corresponds to observables 
of the CS QFT (colored Jones polynomial), can be finally translated 
into an efficient quantum circuit by resorting to a procedure 
similar to that adopted by Aharonov, Jones and Landau in \cite{AhJoLa}.

In section 4.2 we have shown that every unitary matrix 
belonging to the image of representation $\textbf{K}$ is 
efficiently decomposable into a set of universal gates. We now 
prove that the information contained in 
the expectation value $V[L;\textbf{j};q]$ given by
$$
\langle \Phi_{( \textbf{0};\textbf{0} )} ( \hat{l}_1,..., \hat{l}_m^* )|
\textbf{K} \left[b\left( 
\begin{tabular}{ccc}
$\hat{l}_1$ &...& $\hat{l}_m^*$ \\
$\hat{j}_1$ &...& $\hat{j}_m^*$ 
\end{tabular} 
\right)\right]
| \Phi_{( \textbf{0};\textbf{0} )} ( \hat{j}_1,...,\hat{j}_m^* ) \rangle
$$
can be efficiently accessed by a series of measurements.

To begin with, we recall that the standard 
procedure used in quantum computation to evaluate the 
expectation value of a unitary relies on a scheme dubbed 
\textit{Hadamard's trick}. The latter was applied for the 
first time in \cite{AhJoLa} dealing just with the problem 
of evaluating the Jones polynomial. We further recall that 
the notion of approximation used in the present context, 
formalized in \cite{BoFrLo}, is that of \textit{additive approximation}, which has the 
following meaning: given a normalized function $f(x)$, 
where $x$ denotes an instance of the problem in the selected coding, we have an 
additive approximation of its value for each instance $x$ 
if we can associate to $f(x)$ a random variable $Z$ such 
that 
$$
{\rm Pr} \left\{ \left | f(x) - Z \right | \leq \delta \right \} \geq 3/4 \; , 
$$ 
for any $\delta \geq 0$. The time needed to achieve the 
approximation must be polynomial in the size of the problem 
and in ${\delta}^{-1}$. 
The additive characterization of this approximation 
scheme underlies the fact that the interval $\left[ Z-\delta,Z+\delta \right]$, 
which we want to determine, is constructed adding $\pm \delta$ to $Z$. 
It also distinguishes this approximation scheme from the standard 
fully polynomial randomized approximation scheme. 
The normalization adopted for the colored Jones polynomial of a link $L$ is provided by the 
product, over all the link components, of the quantum integer 
related to the dimension of the $SU(2)_q$ irreps labeling the knots. 
The problem we are interested in can 
now be stated as follow.  

\smallskip 

\noindent \textbf{Problem: Approximate colored Jones 
polynomials ($V_L$).} \textit{Given a colored braid $b \in B_{2m}$ of length $\ell$, a 
coloring $\textbf{c}$, a positive integer $k$ and a 
real $\delta > 0$, we want to sample from a random variable $Z$ which 
is an additive approximation of the absolute value of 
the colored Jones polynomial of the plat closure of $b$, evaluated 
at $q = \exp \left ( {\frac{2 \pi i}{k+2}}\right )$, such that the following condition holds true 
$$
{\rm Pr} \left ( \left | V (L; \textbf{j}; q) - Z \right | \leq \delta \right ) \geq 3/4 .
$$} 
\noindent Here the coloring $\textbf{c}$ denotes the 
set of all possibly different irreps of $SU(2)_q$ labeling the component knots of $L$.

In the following we shall provide an 
efficient quantum algorithm for $\textbf{$V_L$}$, which solves it in $O ({\rm poly} (\ell , 
{\delta}^{-1}))$ steps. As in \cite{AhJoLa,WoYa} we 
need the following two lemmas in order to prove the efficiency of the algorithm.

\smallskip 

\noindent \textbf{Lemma 1.} \textit{Given a quantum circuit $U$ 
of length $O ({\rm poly} (n))$, acting on $n$ qubits, and 
given a pure state $| \Phi \rangle$ which can be prepared 
in time $O ({\rm poly} (n))$, then it is possible to sample  
in $O ({\rm poly} (n))$ time from two random variables $a$ and $b$, valued in $\mathbb{Z}_2$, in such 
a way  that $< a + ib > = \langle \Phi | U | \Phi \rangle$.}

\smallskip 

\noindent \textbf{Lemma 2.} \textit{For a sufficiently 
large $N$, given a set of random variables $\{ r_i|i=1,...,N \}$ 
of average value $m$ and square variance $v$}
$$
{\rm Pr} \left( \left | {N}^{-1}\, \sum_{i=1}^{N} r_i - m \right | \geq \delta \right ) \leq 2\, \exp \bigl ( - 
N \delta^2 / (4 v) \bigr).
$$

\smallskip

\noindent The first lemma, which is essentially a 
reformulation of the Hadamard's trick, can be proved as follows. 
Introduce a single-qubit ancilla $\mathfrak{A}$ and denote by $\mathfrak{G}$ 
the Hilbert space of the qubits acted on by $U$. Define the unitary 
$C: \mathfrak{A}\otimes\mathfrak{G} \rightarrow \mathfrak{A}\otimes\mathfrak{G}$ 
through the action:
$$
C \left( |0 \rangle \otimes |\Phi \rangle \right)=|0 \rangle \otimes | \Phi \rangle,
$$
$$
C \left( |1 \rangle \otimes | \Phi \rangle \right)=|1 \rangle \otimes  \left( U|\Phi\rangle \right).
$$
\noindent Initialize then the ancillary qubit in the 
state $\frac{1}{\sqrt{2}} \left( |0\rangle + |1\rangle \right) 
\in \mathfrak{A}$ and prepare the system in the initial state $| 
\Phi \rangle$. The action of $C$ maps the initial state into 
$|\Psi\rangle \in \mathfrak{A}\otimes\mathfrak{G}$
$$
|\Psi\rangle \equiv \frac{1}{\sqrt{2}} \left( |0\rangle 
|\Phi\rangle + |1\rangle \left( U|\Phi\rangle \right) \right).
$$
\noindent The reduced density matrix $\rho^{{\mathfrak A}}$ of the ancilla is thus equal to \\
$$
\rho^{{\mathfrak A}}={\rm Tr}_{{\mathfrak G}} |\Psi\rangle \langle \Psi|= \frac{1}{2}{\rm Tr}_{{\mathfrak G}}
\left(
\begin{tabular}{cc}
$\Phi$ & $\Phi U^{\dagger}$ \\
$U \Phi$ & $U \Phi U^{\dagger}$
\end{tabular}
\right)=
\frac{1}{2}
\left(
\begin{tabular}{cc}
$1$ & $\langle \Phi | U^{\dagger} | \Phi\rangle$ \\
$\langle \Phi | U | \Phi\rangle$ & $1$
\end{tabular}
\right)=
$$

$$
=\frac{1}{2} \left( \mathbb{I}_2+\sigma_x {\rm{Re}}
\langle \Phi |U|\Phi \rangle + \sigma_y {\rm{Im}} \langle \Phi |U|\Phi\rangle \right),
$$
\noindent where $\Phi$ denotes the density matrix 
$ |\Phi \rangle \langle \Phi| $ and $\sigma_x, \sigma_y$ are Pauli matrices.

The mean value of a sequence of measurements 
of $\sigma_x$ will approach ${\rm{Re}} \langle \Phi |U|\Phi \rangle$, 
whereas the mean value of a sequence of measurements of $\sigma_y$ 
will approach ${\rm{Im}} \langle \Phi |U|\Phi\rangle$.\\
The second lemma, which is a modified version of the well 
known Chernoff bound, ensures us that we can approximate 
these values polynomially in the number $N$ of samplings and 
in the inverse of the precision $\delta^{-1}$.

\begin{figure}[htbp]
	\centering
		\includegraphics[height=2cm]{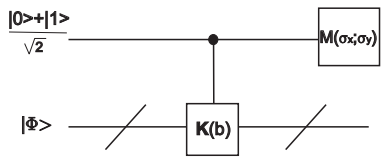} 
        \caption{\small{The circuit realizing the Hadamard's trick for the Kaul representation of the braid $b$. $M(\sigma_x ; \sigma_y)$ denotes quantum measurement of either $\sigma_x$ or $\sigma_y$.}}
	\label{fig:hadamard}
\end{figure}

\noindent Summarizing the qubit model for the Kaul representation can be used to efficiently compile 
a unitary representation of the colored braid group, 
and a sampling procedure can then be used to efficiently estimate 
the value of the colored Jones polynomial. The circuit 
that realizes all these steps is schematically depicted in fig.[\ref{fig:hadamard}].

\noindent In conclusion, the sampling lemma tells us 
that measurements of $\sigma_{x}$ on the first qubit will provide the value 
for ${\rm{Re}} \left ( V \left ( L,\textbf{j}, q \right )\right )$, 
while measurements of $\sigma_{y}$ on the 
first qubit will provide the value for ${\rm{Im}} \left 
( V \left ( L,\textbf{j},q \right )\right )$.

\section{Conclusions}
The $q$-deformed spin network model provides the natural setting 
for a quantum automaton capable of processing the braid group language. 
Coding of information in the spin network is done in the frame of the coupling scheme 
associated with the 'co-power' $\Delta^n (SU(2)_q)$ (iterated co-product) 
of the network $q$-algebra. Such $\textit{parenthesized}$ coding lends itself 
quite naturally to deal with a number of hard combinatorial problems, ranging 
from finite groups word or isomorphism problems \cite{Deh, Bar} to the 
evaluation of topological invariants. Work is in progress on problems of the 
former type. Focusing mainly on the latter, 
for consistency with our introductory physical setting, we discuss here briefly 
the role played by the colored link polynomial introduced above 
in $3$--dimensional geometric topology \cite{Oht,MeTh} 
in view of Thurston's `geometrization programme' \cite{Thu}.
Indeed the possible extension of our quantum algorithm to address the 
(classically computationally hard) problems
outlined below would represent a major breakthrough both
in quantum computation and in the theory of closed (hyperbolic)
$3$--manifolds. On the physical side, such an achievement would
open the possibility of `controlling' the quantum algorithmic complexity
of three dimensional quantum gravity models. It is worth recalling 
here the alternative point of view introduced by the recent attempt by 
S. Lloyd of unifying quantum mechanics and gravity; where the very geometry of space-time 
is a construct derived from the underlying quantum computation \cite{Llo}.

\subsection{Reshetikhin--Turaev quantum invariants of $3$--manifolds 
and their quantum complexity}
At its foundation, knot theory
is a branch of geometric topology, since it allows us to explore
$3$--dimensional spaces by `knotting' phenomena, namely embedded
knots `interact' with the topological structure of the ambient $3$--manifold
$\mathcal{M}^3$. The content of the latter remark is made  
more stringent by a theorem which asserts that every closed connected
orientable $3$--manifold can be obtained by Dehn `surgery' along a framed link 
embedded in the $3$--sphere $S^3$ (we refer to \cite{Lic} for definitions and proofs).
Roughly speaking, a tubular neighborhood of each component of the embedded link $L$,
represented by $S^1 \times D^2$ ($D^2$ being the $2$--disk), is removed and
replaced by  $D^2 \times S^1$ in a suitable way, generating the new manifold.
Formally
\begin{equation}\label{surg}
(S^3,\,L) \;  \longrightarrow \;\mathcal{M}^3_{L}\,  \doteq S^3 \setminus L.
\end{equation}
It can also be shown that equivalent links, namely links which are ambient isotopic,
give rise to the same type of $3$--manifolds (the manifolds obtained by surgeries
in the $3$--sphere
along equivalent links are homeomorphic).

The idea that the Jones polynomial at a root of unity $q$ can be 'amplified' to achieve
a $3$--manifold quantum invariant dates back to Witten, and was further implemented
by a number of authors (\cite{Lic} and references therein). Such invariants, which correspond
to the partition function (\ref{CSfunct}) evaluated for a manifold $\mathcal{M}^3_{L}$,
are linear sums of Jones polynomials of copies of the link with the components replaced
by various parallels of the original components. The authors of \cite{KaLo} propose 
to address the problem of designing quantum algorithms for 
Witten invariants by resorting to
Temperley--Lieb algebra techniques.

The quantum algorithm for the colored Jones polynomials discussed in 
section 4, allows us to conjecture that the associated colored $3$--manifold 
quantum invariants at a fixed root of unity can be actually evaluated in a quite  straightforward way. 
The explicit expression of the (Witten--)Reshetikhin--Turaev quantum invariant
for a $3$--manifold $\mathcal{M}^3_{L}$
to be used for computational purposes 
was proposed by Kirby and Melvin \cite{KiMe} and reads 
$$
\tau ({\cal M}^{3}_{L};q) = \alpha_{L} \sum_{\textbf{j}} \, [\textbf{j}]\,
{\cal E}_{j_1...j_s} [L;q]\,,
$$
 where $\mathbf{j}$ stands for the collective assignment of colorings to the link components, 
the summation is over all admissible colorings  and
$[\mathbf{j}] = \prod_{i=1}^{s}[2j_i+1]$. 
$\mathcal{E}_{j_1...j_s}\,[L;q]$ is given in (\ref{Wilexpect}) and
$\alpha_{L}$ $=b^{n_L}\,c^{\sigma_{L}}$. Here $b$ and $c$ are numbers depending on
the integer $k$ ($b=\sqrt{(2/k )}\, \sin \frac{\pi}{k}$ and $c=\exp [-2\pi i(k-2)/8k]$),
$n_L$ is the number of link components and $\sigma_L$ is the signature of the linking matrix
of $L$. The linking matrix $M_L$ of a framed link $L$ 
is a symmetric matrix whose entry $(M_L)_{ij}$ for $i\neq j$ 
is the linking number between components $i$ and $j$ of $L$. The diagonal elements of
$M_L$ are defined to be the integers that give the framing of the individual components.
The linking matrix , defined here in combinatorial terms, is related to the topology of
$\mathcal{M}^3_{L}$ because its determinant (if it is non--zero) is the order of the
first homology group of the manifolds.

\subsection{The volume conjecture for hyperbolic $3$--manifolds}
In the framework of Thurston's geometrization program \cite{Thu}
(all three dimensional manifolds can be reconstructed starting from
eight types of model geometries), hyperbolic $3$--manifolds play a special role.
Recall that a $3$--manifold is hyperbolic if it is endowed with a complete 
riemannian metric with constant negative sectional curvature. 
The most interesting case is that of complete hyperbolic manifolds with finite
volume (the volume being evaluated in the given metric) since the Mostow 
rigidity theorem asserts that any two such manifolds are homeomorphic if and
only if they are homotopically equivalent (we refer to \cite{FoKu} for details 
and original references). Typical instances of such manifolds are obtained as
quotients of the hyperbolic $3$--space $\mathbb{H}^3$ by discrete subgroups of 
the full isometry group of $\mathbb{H}^3$.

Consider the set of all complete hyperbolic 
$3$--manifolds with a finite volume. Then  
the set of volumes is totally ordered; moreover there exist only finitely many different hyperbolic 
$3$--manifolds with the same volume \cite{Gro}.
Thus the volume of an hyperbolic manifold (unlike what happens in the euclidean and
elliptic cases) can be considered as a topological invariant. Computer geometry 
is the  branch of geometric topology devoted to the calculation of these invariants:
it still exhibits many open problems interesting 
for a quantum-computational approach but it is most intriguing that such a hard `computational'
approach raised discussions among mathematicians 
about the ``philosophical'' question of the effectiveness  
of mathematical proofs.

At first sight the above remarks on hyperbolic 
volumes does not seem related to our central issue of 
colored quantum invariant of $3$--manifolds. However this is not the case: 
a connection can be easily recognized by 
observing that most manifolds obtained by surgery on framed knots (links) in the
$3$--sphere can be endowed with hyperbolic metrics. 
Let us focus for example on `hyperbolic knots', namely 
those knots which give rise by surgery to (finite volume) 
hyperbolic $3$--manifolds: the `volume conjecture' proposed 
by \cite{Kas,MuMu} (see also the review \cite{Oht}
for extended versions) can be cast in this case in the form
\begin{equation}\label{volcon}
2\pi\,\lim_{N \rightarrow \infty}\;\frac{\log |\mathit{J}_N (K)|}{N}\;=\;
\mbox{Vol} \,(S^3 \setminus K)\,,
\end{equation}
 where $K$ is a hyperbolic knot and the notation $\mathit{J}_N (K)$
 stands for the $N$--colored  polynomial of $K$ evaluated at $q=\exp (2\pi i/N)$.

Notice that all the quantum algorithms dealing with link polynomials
 are established for a fixed choice of the root of unity $q$ appearing in the argument
 of the invariants, while the volume
conjecture involves the analysis of the asymptotic behavior of `single-colored'
polynomial of the same knot for increasing values of the coloring itself. 
Recently, Aharonov and Arad \cite{AhAr} have addressed an asymptotic analysis ($k \rightarrow \infty$)
 for the Jones polynomial, still $\frac{1}{2}$--colored. 
 It would be interesting to explore the possibility of
borrowing some of their techniques to test the conjecture (\ref{volcon})
within the computational framework designed for colored polynomials.  

\smallskip

\noindent \textbf{Acknowledgments}\\
We are in debt with Romesh Kaul for clarifying remarks on his work on colored polynomials.


\smallskip

\noindent \textbf{References}

\begin{thebibliography}{99}

\bibitem{Jon}
Jones V F R 1985
{\it Bull. Amer. Math. Soc.} {\bf 12} 103

\bibitem{AhJoLa}
Aharonov D, Jones V and Landau Z 2005 {\rm A polynomial 
quantum algorithm for approximating the Jones polynomial} \textit{Preprint}: quant-ph/0511096

\bibitem{GaMaRa1} Garnerone S Marzuoli A and Rasetti A 
\rm{Quantum automata, braid group and link polynomials} \textit{Preprint}: quant-ph/0601169

\bibitem{ArDeMi}
Arnowitt R, Deser S and Misner C W 1962 
{\rm The dynamics of general relativity} 
{\it Gravitation: An Introduction to Current Research}
ed L Witten (New York: Wiley)

\bibitem{AdBaSc}
Adler R, Bazin M and Schiffer M 1965
{\it Introduction to General Relativity}
(New York: McGraw--Hill)

\bibitem{MiThWh}
Misner C W, Thorne K S and Wheeler J A 1973
{\it Gravitation} (San Francisco: W H Freeman and Co.)

\bibitem{Whe}
Wheeler J A 1968
{\rm Superspace and the nature of quantum geometrodynamics}
{\it Battelle Rencontres' ed C M DeWitt and J A Wheeler}
(New York: W A Benjamin)

\bibitem{Wei}
Weinberg S 1995 {\it The Quantum Theory of Fields} (Vol. 1: Foundations)
(Cambridge: Cambridge University Press)

\bibitem{HaHa}
Hartle J B and Hawking S W 1983
{\it Phys. Rev.} D {\bf 28} 2960

\bibitem{Reg}
Regge T 1961
{\it Nuovo Cimento} {\bf 19} 558

\bibitem{AmCaMa}
Ambjorn J, Carfora M and Marzuoli A 1997
{\it The Geometry of Dynamical Triangulations}
Lect. Notes in Physics {\bf m 50}
(Berlin: Springer--Verlag)

\bibitem{AmDuJo}
Ambjorn J, Durhuus B and Jonsson T 1997
{\it Quantum Geometry}
(Cambridge: Cambridge University Press)

\bibitem{YaMi}
Yang C N and Mills R 1954
{\it Phys. Rev.} {\bf 96} 191

\bibitem{Jac}
Jackiw R 1980
{\it Rev. Mod. Phys.} {\bf 52} 661

\bibitem{EgHaGi}
Eguchi T, Gilkey P B and Hanson A J 1980 {\it Phys. Rep.} \textbf{66} 213

\bibitem{Ati}
 Atiyah M F 1989 {\rm Topological quantum field theories}
{\it Publ. Math. IHES} {\bf 68} 175

\bibitem{BiBlRa}
Birmingham D, Blau M, Rakowski M and Thomson G 1991
{\it Phys. Rep.} {\bf 209} 129

\bibitem{Qui}
Quinn F 1995
{\rm Lectures on axiomatic topological quantum 
field theories} ed D S Freed {\it et al. Geometry and Quantum 
Field Theory} (Providence RI: Amer. Math. Soc.)

\bibitem{Wit}
Witten E 1989 
{\it Commun. Math. Phys.} {\bf 121} 351

\bibitem{Car}
Carlip S 1998
{\it Quantum Gravity in 2+1 dimensions}
(Cambridge: Cambridge University Press)

\bibitem{ReTu}
Reshetikhin N and Turaev V G  1991
{\it Invent. Math.} {\bf  103} 547  

\bibitem{KiMe}
Kirby R and Melvin P 1991 
{\it Invent. Math.} {\bf 105} 473

\bibitem{Lic}
Lickorish W B R 1997
{\it An Introduction to Knot theory}
(New York: Springer)

\bibitem{Gua}
Guadagnini E 1993 {\it The Link Invariants of the Chern--Simons
Field theory} (Berlin: de Gruyter)

\bibitem{RaGoKa}
Ramadevi P, Govindarajan T R and Kaul R K  1994 
{\it Mod.  Phys. Lett.} A {\bf 9} 3205

\bibitem{GMRlaser}
Garnerone S, Marzuoli A and Rasetti M 2006
{\rm Quantum Knitting}
{\it Preprint}: quant-ph/ 0606137

\bibitem{Kau}
Kaul R K 1994 {\it Commun. Math. Phys.} {\bf 162} 289

\bibitem{Cho} 
Chomsky N 1956 {\it IRE Transactions on Information Theory} {\bf 2} 113

\bibitem{HoUl} Hopcroft J E and Ullman J D 1979 {\it Introduction to 
Automata Theory, Languages and Computation}, (Reading MA: Addison--Wesley).

\bibitem{MoCr}
Moore C and Crutchfield J P 2000 {\it Theor. Comput. Sci.} {\bf 37} 275

\bibitem{MaRa1}
Marzuoli A and Rasetti M 2002 {\it Phys. Lett.} A {\bf 306} 79

\bibitem{MaRa2} 
Marzuoli A and Rasetti M 2005 {\it Ann. Phys.} {\bf 318} 345

\bibitem{BiBr}
Birman J S and Brendle T E 2004 {\rm Braids: a survey} {\it Preprint}: math/ 0409205

\bibitem {Bir}
Birman J S 1974 {\it Braids, Links and Mapping Class Groups} (Princeton NJ: Princeton Univ. Press)

\bibitem{HaReCh}
Harrow A, Recht B and Chuang I L 2002 {\it Journal of Mathematical Physics} \textbf{43} 4445

\bibitem{BoFrLo} 
Bordewich M, Freedman M, Lovasz L and Welsh D 2006
{\rm Approximate counting and quantum computation} to appear in \textit{Combinatorics, Probability and Computing}

\bibitem{WoYa} 
Wocjan P and Yard J 
{\rm The Jones polynomial: quantum algorithms and applications in quantum complexity theory} {\it Preprint}: quant-ph/0603069

\bibitem{Deh}
Dehornoy P 
{\rm The group of parenthesized braids} 
{\it Preprint}: math/0407097

\bibitem{Bar}
Bar-Natan D 1997
{\rm{Non-associative tangles}} \textit{Geometric Topology} \rm{ed W H Kazez} {\it et al.} (Providence R I: Amer. Math. Soc. and International Press) \rm{pp 139-183}

\bibitem{Oht} 
Ohtsuki T ed 2002 {\it Problems on invariants of knots and
3--manifolds} RIMS Geometry and Topology Monographs, Vol. 4

\bibitem{MeTh}
Menasco W and Thistlethwaite eds 2005
{\it Handbook of Knot Theory} (Amsterdam: Elsevier)

\bibitem {Thu}
Thurston W P 1997 {\it Three--Dimensional Geometry and Topology}
(Vol. 1) (Princeton: Princeton University Press)

\bibitem{Llo}
Lloyd S 2005 {\rm A Theory of quantum gravity based on quantum computation} 
{\it Preprint}:quant-ph/0501135

\bibitem{KaLo}
Kauffman L H and Lomonaco S J
{\rm q-deformed spin networks, knot polynomials and anyonic topological quantum computation}
{\it Preprint}:quant-ph/0606114

\bibitem{FoKu}
Fomenko A T and Kunii T L 1997
{\it Topological Modeling for Visualization}
(Tokyo: Springer-Verlag)

\bibitem{Gro}
Gromov M 1981
{\it Hyperbolic manifolds} Lect. Notes in Math \textbf{842} 40 (Berlin: Springer-Verlag)

\bibitem{Kas}
Kashaev R M  1995 {\it Mod. Phys. Lett.} A {\bf 10} 1409

\bibitem{MuMu}
Murakami H and Murakami J 2001
{\it Acta Math.} {\bf 186} 85

\bibitem{AhAr}
Aharonov D and Arad I
{\rm The BPQ--hardness of approximating the Jones polynomial}
{\it Preprint}:quant-ph/0605181

\end{thebibliography}
\end{document}